\documentstyle[preprint,tighten,aps,epsfig]{revtex}

\begin{document}
\title{Screened {\bf potential} and the baryon spectrum}
\author{J. Vijande$^{\, 1}$, P. Gonz\'alez$^{\, 2}$, H. Garcilazo$^{\, 3}$, and A.
Valcarce$^{\, 1}$}
\address{$^{\, 1}$ Grupo de F\' \i sica Nuclear \\
Universidad de Salamanca, E-37008 Salamanca, Spain}
\address{$^{\, 2}$ Dpto. de F\' \i sica Te\'orica and IFIC\\
Universidad de Valencia - CSIC, E-46100 Burjassot, Valencia, Spain}
\address{$^{\, 3}$ Escuela Superior de F\' \i sica y Matem\'aticas \\
Instituto Polit\'ecnico Nacional, Edificio 9,\\
07738 M\'exico D.F., Mexico}
\maketitle

\begin{abstract}
We show that in a quark model scheme the use of a screened potential,
suggested by lattice QCD, instead of an infinitely rising one with the
interquark distance, provides a more adequate description of the high-energy
baryon spectrum. In particular an almost perfect parallelism between the
predicted and observed number of states comes out throwing new light about
the so-called missing resonance problem.

\vspace*{2cm} \noindent Keywords: nonrelativistic quark models, baryon
spectrum, missing states \newline
\newline
\noindent Pacs: 12.39.Jh, 14.20.-c, 14.20.Gk
\end{abstract}

\newpage

\section{Introduction}

Constituent quark models of baryon structure are based on the assumption of
effective quark degrees of freedom so that a baryon is a three-quark
color-singlet state. Quarks are bound by means of a phenomenological
$qq$ potential attempting to incorporate observed features of the
hadron spectra. Although not rigorously proven from QCD,
numerical simulations for heavy quarks in the lattice in the quenched
approximation (considering only valence quarks) show out a $qq$ confining
potential linearly rising with the interquark distance \cite{BALI}. This
potential produces an infinite discrete hadron spectrum. The
implementation of a confining force of this type with effective one gluon
and/or boson exchanges, or other effective interactions, turns out to be
fruitful in the construction of quark potential models that provide with a
precise description of baryon spectroscopy \cite{ISGU,DESP,VAGO,GLOZ}.
However an outstanding problem remains unsolved: all models predict a
proliferation of baryon states at excitation energies above 1 GeV which are
not experimentally observed as resonances. This difference between the quark
model prediction and the data about the number of physical resonances is
known as the missing resonance problem. An explanation for such a
discrepancy has been given by realizing that most (but not all) of the
missing states could be too inelastic to be easily observed \cite{KONI,CAPS}.
To test experimentally the situation is a current objective of the CLAS
collaboration in TJNAF \cite{RIPA} and also the BES collaboration in Beijing 
\cite{ZOUB}.

In this article we propose to revisit the baryon spectrum centering the
attention in its high-energy excited states. As this sector in spectroscopic 
quark models is essentially determined by the linear confinement, we
are driven to refine the potential. To do this we incorporate effects
associated to the creation of light $q\bar q$ pairs out of vacuum. 
For heavy quarks the spontaneous pair creation between them may give
rise to a breakup of the color flux tube in such a way that the quark-quark
potential does not rise with the interquark distance but it reaches a
maximum saturation value. Though from previous lattice calculations a
potential parametrization of these screening effects was proposed \cite{BORN}
there is not a convincing confirmation of string breaking \cite{CONF}.
However a quite rapid cross-over from a linear to a flat potential is well
established in SU(2) Yang-Mills theory \cite{SUYM}. We shall assume
hereforth that such screening occurs in QCD and can be approximately
parametrized, even for light quarks, through a potential as proposed
in Ref. \cite{BORN}. Such an assumption has been employed in the literature
to study the nonstrange baryon spectrum \cite{CHINO}, the heavy meson
spectrum \cite{HEME}, and as a possible explanation of the nonlinear
hadronic Regge trajectories \cite{GOLD}. For the baryon spectrum an
improvement in the description of the higher energy excitations with respect
to the linear confining potential case was obtained. For the heavy mesons a
pretty good description of bottomonium was accomplished. Nevertheless the
more striking predictions associated to screening, say the breaking of the
color flux tube between quarks (or quark and antiquark) and the
corresponding finiteness of the bound state spectrum could not be tested in
the meson sector because of the lack of sufficient data at the high-energy
excitation region. On the other hand in Ref. \cite{CHINO} the results for
the baryon sector, obtained by means of a variational method with gaussian
trial wave functions, were restricted to $J\leq 7/2$ and no threshold
stability analysis was carried out. Our hope here is that the more extensive
and deep analysis of the known nonstrange high-energy baryon spectrum may
allow us to establish, or at least make feasible, the validity of such
predictions.

To this purpose we start in Sec. II by revising the nonstrange baryon
spectrum obtained with a minimal model that includes an effective
linear confinement plus a one-gluon-exchange (OGE) like potential. From it
we establish the connected missing resonance problem. Then in Sec. III we
introduce screening effects in the potential and derive the value of the
parameters by plausibility arguments and fitting to data. Sec. IV is devoted
to the results we obtain for the whole spectrum and their comparison to
data. Finally in Sec. V we summarize our work and resume our main
conclusions.

\section{Strict confinement}

Lattice calculations in the quenched approximation derive, for heavy
quarks, a confining interaction linearly dependent on the interquark
distance \cite{BALI}. This form of strict confinement has been widely
used for light quarks when studying the baryon spectrum within a quark
model framework \cite{ISGU,DESP,VAGO,GLOZ}. To illustrate our discussion and
for the sake of simplicity we shall make use of a nonrelativistic quark
model potential containing besides the linear confinement a minimal OGE term
(Coulomb and spin-spin). Such a model was proposed in Ref. \cite{BACO} to
describe the meson spectrum and it was later on applied to the baryon case 
\cite{BRAC}. The $qq$ potential reads (there is a factor of difference $1/2$
with respect to the $q\overline{q}$ case, coming from the assumed $\vec{%
\lambda}_{i}\cdot \vec{\lambda}_{j}$ color structure): 
\begin{equation}
V(r_{ij})={\frac{1}{2}}\left[ -{\frac{\kappa }{r_{ij}}}+{\frac{r_{ij}}{a^{2}}%
}+{\frac{{\hbar ^{2}\kappa _{\sigma }}}{{m_{i}m_{j}c^{2}}}}{\frac{{%
e^{-r_{ij}/r_{0}}}}{{r_{0}^{2}r_{ij}}}(}\vec{\sigma}_{i}\cdot \vec{\sigma}%
_{j})\right]  \label{bcn}
\end{equation}
where $r_{ij}$ is the interquark distance, $m_{i}$ the mass of quark $i$, $%
\vec{\sigma}$ the Pauli matrices and $\kappa $, $a$ and ${\kappa _{\sigma }}$
are constants. The hyperfine (spin-spin) interaction has been regularized in
order to avoid the unbound spectrum that would cause a contact term,
$\delta (\vec{r}_{ij})$, as obtained from the nonrelativistic reduction
of the OGE diagram in QCD. The values of the parameters appear 
in Table \ref{t1}.

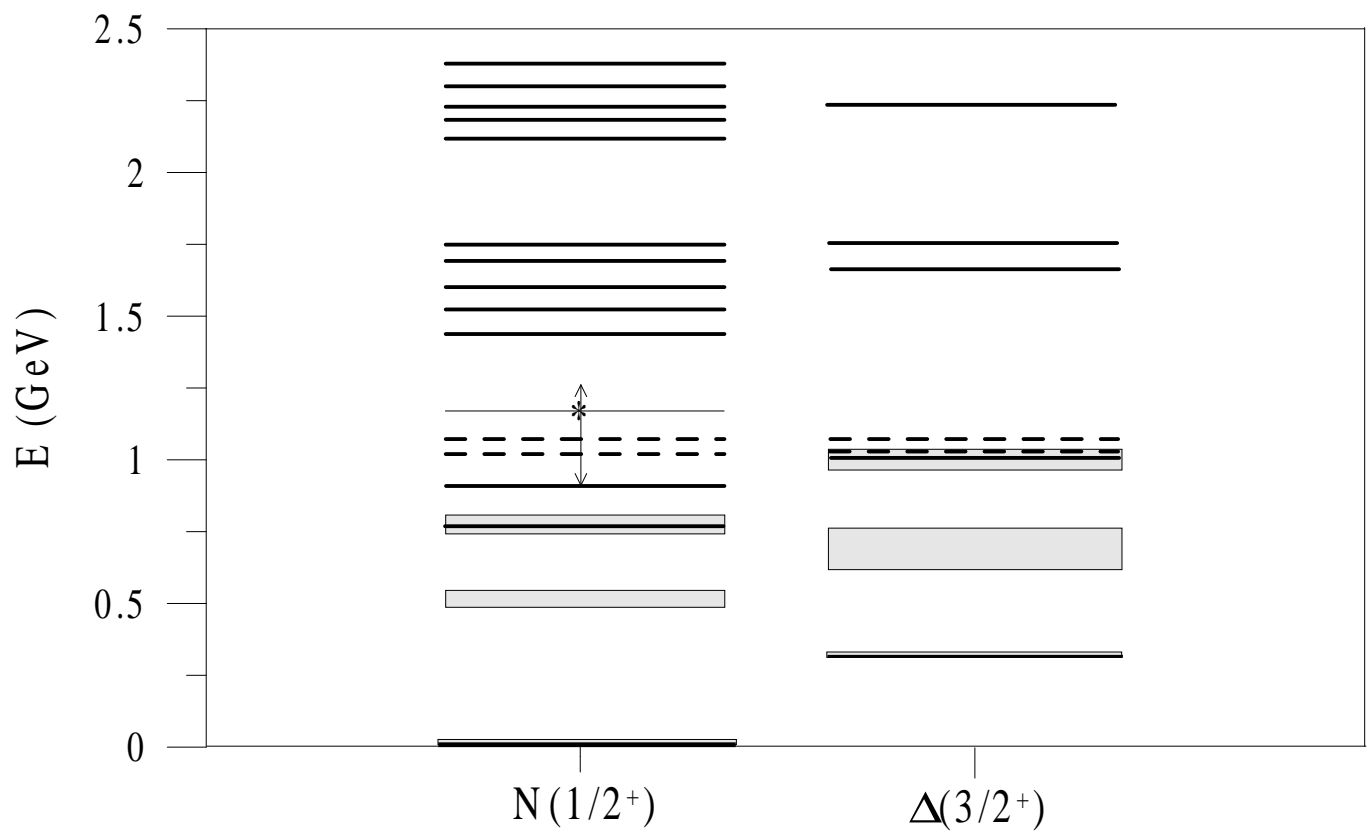
\begin{figure}[tbp]
\caption{ Relative energy $N(1/2^{+})$ and $\Delta (3/2^{+})$ spectrum up to
2.5 GeV excitation energy for the potential of Ref. \protect\cite{BRAC}. The
solid and dashed lines correspond to the Faddeev results including $\ell $
and $\protect\lambda $ up to 5, and for $L<3$ as explained in the text. The
shaded regions, whose size stands for the experimental uncertainty,
represent three or four star resonances in the notation of Ref. \protect\cite
{HAGI}. The energy of the one star resonance in the $N(1/2^+)$ case is
denoted by a thin solid line with one star over it and its experimental
uncertainty by a vertical line with arrows.}
\label{f1}
\end{figure}

Fig. \ref{f1} shows a part [for $N(1/2^{+})$ and $\Delta (3/2^{+})$] of the
predicted nonstrange baryon spectrum and its comparison to data. We have
plotted by the solid line the $L=0$ multiplets ($L$ is the total orbital
angular momentum) and by the dashed line the first two excited states with $%
L\neq 0$. As clearly seen two major problems show up. On the one hand the
position of Roper resonances (first positive parity excitations) for $N$\
and $\Delta $, on the other hand the proliferation of baryon states above 1
GeV excitation energy that have not been experimentally seen. The first
problem, say the location in energy of the Roper resonances, has two
different aspects, one is the high excitation energy predicted that comes in
part from the large strength used for the Coulomb interaction, the
other is its inverted position (with respect to data) relative to the first
negative parity states. This problem was long-ago related to relativistic 
corrections \cite{CARL}. Since then many different solutions have been
proposed \cite{DESP,GLOZ,STST,BAJE,METS,KRSP,VAL1,FURU}. For the second
problem, say the theoretically predicted excess of high energy excitations
(missing state problem), the best known proposed solution is based on the
weak pion coupling that most of these predicted excited states may have what
would make them very difficult to be detected experimentally \cite
{KONI,CAPS,CAPS2,KONIB,STAN}.

\section{Screened potential}

The consideration in the lattice of sea quarks apart from valence quarks
(unquenched approximation) suggests a screening effect on the potential when
increasing the interquark distance \cite{BALI}. Creation of light $q%
\overline{q}$ pairs out of vacuum in between the quarks becomes
energetically preferable resulting in a complete screening of quark color
charges at large distances (in an adiabatic approach this could be
interpreted as if the confining energy ceases to increase because it is also
employed in the pair creation). Then the breaking of color flux tubes, or
the splitting of the quarks, may occur. Beyond the splitting energy the same
interaction with the sea quarks will give rise to hadronization. Otherwise
said the decay process takes place.

In the 80's a specific parametrization of these effects was given in the
form of a screening multiplicative factor in the potential reading $\left[
\left( 1-e^{-\mu r_{ij}}\right) /\mu r_{ij}\right] $ where $\mu$ is 
a screening parameter \cite{BORN}. As a consequence the
hadron spectrum becomes finite. Let us realize that for sufficiently
larger distances the screened linear term in the potential becomes
the dominant one. Thus the highest excited states are presumably
``confinement states'', i. e. they may be determined by considering only the 
screened linear interaction. The other way around, from the
experimentally detected higher excited states one might infer approximate
values of the screened linear potential parameters. Unfortunately for
the heavy meson spectrum high energy excitation data are very poor and we
cannot pursue the phenomenological analysis in this manner. On the contrary,
if we assume the same form of screening for light quarks, the
nonstrange baryon spectrum seems to be promising to this respect. For
the sake of simplicity we will restrict ourselves to a screened Bhaduri-type
potential. Following the notation employed in Eq. (\ref{bcn}), we
write the potential: 
\begin{equation}
V_{b}(r_{ij})=\frac{1}{2}\left[ \frac{r_{ij}}{\overline{a}^{2}}-\frac{%
\overline{\kappa }}{r_{ij}}+\frac{\hbar ^{2}\overline{\kappa _{\sigma }}}{%
m_{i}m_{j}c^{2}}\frac{e^{-r_{ij}/\overline{r_{0}}}}{\overline{r_{0}}%
^{2}r_{ij}}(\vec{\sigma _{i}}\cdot \vec{\sigma _{j}})\right] \left( \frac{%
1-e^{-\mu _{b}\,\,r_{ij}}}{\mu _{b}\,\,r_{ij}}\right)
\end{equation}
where the hat over the parameters indicates that they are different than in
the nonscreened case.

Notice that there will be a splitting energy, equal or bigger than $3/(2%
\overline{a}^{2}\mu _{b})$ (since there are three pairs of quarks), for the 
$3q$ system to be split into three free quarks. But equally important in this
case in order to analyze the stability of the system is the calculation of
the energy thresholds for only one quark to be released. For $3q$ binding
energies higher than these thresholds a ($2q$ state $+1q$ free state) will
be a more probable configuration that, through hadronization, will decay
into a baryon and a meson or multimeson states. From now on we shall refer
to these energy thresholds as $1q$ ionization thresholds. Let us note that
as these energy thresholds may be well below the top of the $3q$ effective
potential, they do not completely prevent the possible existence of
metastable $3q$ bound states at higher energy assumed that for dynamical
reasons the induced decays are suppressed.

Concerning the value of the parameters we can give first some arguments of
plausibility before going to a more refined fit from data. As the screening
multiplicative factor reduces to 1 when $r_{ij}\rightarrow 0$ the potential
comes dominated in this limit by the OGE piece with exactly the same form
than in the nonscreened case. So we expect the parameters of the OGE
interaction, $\overline{\kappa },\overline{\kappa _{\sigma }}$ and 
$\overline{r_{0}}$ to be quite close to the corresponding values of $\kappa$,
$\kappa _{\sigma }$ and $r_{0}$ given in Sec. II in order to get a similar
description of the low-lying baryon spectrum. However as commented above the
large Coulomb strength in the nonscreened case is to a good extent
responsible for the incorrect position predicted for the radial excitations.
So we shall reduce $\overline{\kappa }$ against $\kappa $ as much as the
preservation of the quality of the predicted spectrum allows. This has the
bonus of reducing the relative importance of the Coulomb interaction 
versus the linear one what is very convenient for a fit of the screened
linear potential parameters from the highest energy excitation data.

With respect to $1/(2\overline{a}^{2})$ we expect its value to be
bigger than $1/(2a^{2})=$ 470.5 MeV fm$^{-1}$ from Sec. II. In this manner
the linear potential strength reduction coming from the screening can
be compensated and the quality of the medium-lying baryon spectrum obtained
in Ref. \cite{BRAC}, for which the linear term was already playing some
role, not spoiled. Regarding $\mu _{b}$ we expect it to be, in the light quark
case, a completely effective parameter.

In order to be more precise we center our attention in the experimental
higher energy resonances: there are a $N(7/2^{-})$ state at 2190 MeV, a $%
N(9/2^{+})$ at 2220 MeV, a $N(9/2^{-})$ at 2250 MeV and a $\Delta (11/2^{+})$
at 2420 MeV cataloged as well established $(\ast \ast \ast \ast )$ states by
the Particle Data Group \cite{HAGI}. For the nucleon there is also a very
likely $N(11/2^{-})$$(\ast \ast \ast )$ state at 2600 MeV and a fair
evidence of a $N(13/2^{+})$$(\ast \ast )$ at 2700 MeV. For the $\Delta $
there are two more uncertain states, the $\Delta (13/2^{-})$$(\ast \ast )$
at 2750 MeV and the $\Delta (15/2^{+})$$(\ast \ast )$ at 2950 MeV. It turns
out very difficult to reasonably accommodate within the experimental errors
(also more uncertain for resonances above 2500 MeV) and with the same 
screened linear potential the well established resonances and the uncertain
ones altogether. Assuming that this may be an indication of the presence of
the continuum we shall restrict ourselves to the highest $(\ast \ast \ast
\ast )$ particles, that we shall consider to be close to threshold, to fix
the screened linear potential parameters.

\section{Results}

To get the nonstrange baryon spectrum we proceed to solve the
Schr\"{o}dinger equation by two different procedures: i) the hyperspherical
harmonic (HA) expansion method \cite{GIUS} and ii) the Faddeev method. The
HA treatment allows a more intuitive understanding of the wave functions in
terms of the hyperradius of the whole system. As a counterpart one has to go
to a very high order in the expansion to get convergence. To assure this we
shall expand up to $K=24$ ($K$ being the great orbital determining the order
of the expansion). In the Faddeev calculation in order to assure convergence
we shall include ($l,\lambda ,s,t$) configurations ($l$ is the orbital
angular momentum of a $2q$ pair, $\lambda $ is the orbital angular momentum
of the third quark with respect to the center of mass of the $2q$, and $s$
and $t$ are the spin and isospin of the $2q$ respectively) up to $l$=5 and $%
\lambda $=5 \cite{VAGO}. Differences in the results for the $3q$ bound state
energies obtained by means of the two methods turn out to be at most of 5
MeV. Let us realize that we can calculate with precision with any method
only below or close above (by a continuity procedure) the $1q$ ionization
thresholds. Higher in energy the oscillatory behavior corresponding to the $%
1q$ free state introduces numerical uncertainties and the level of
convergence is lost.

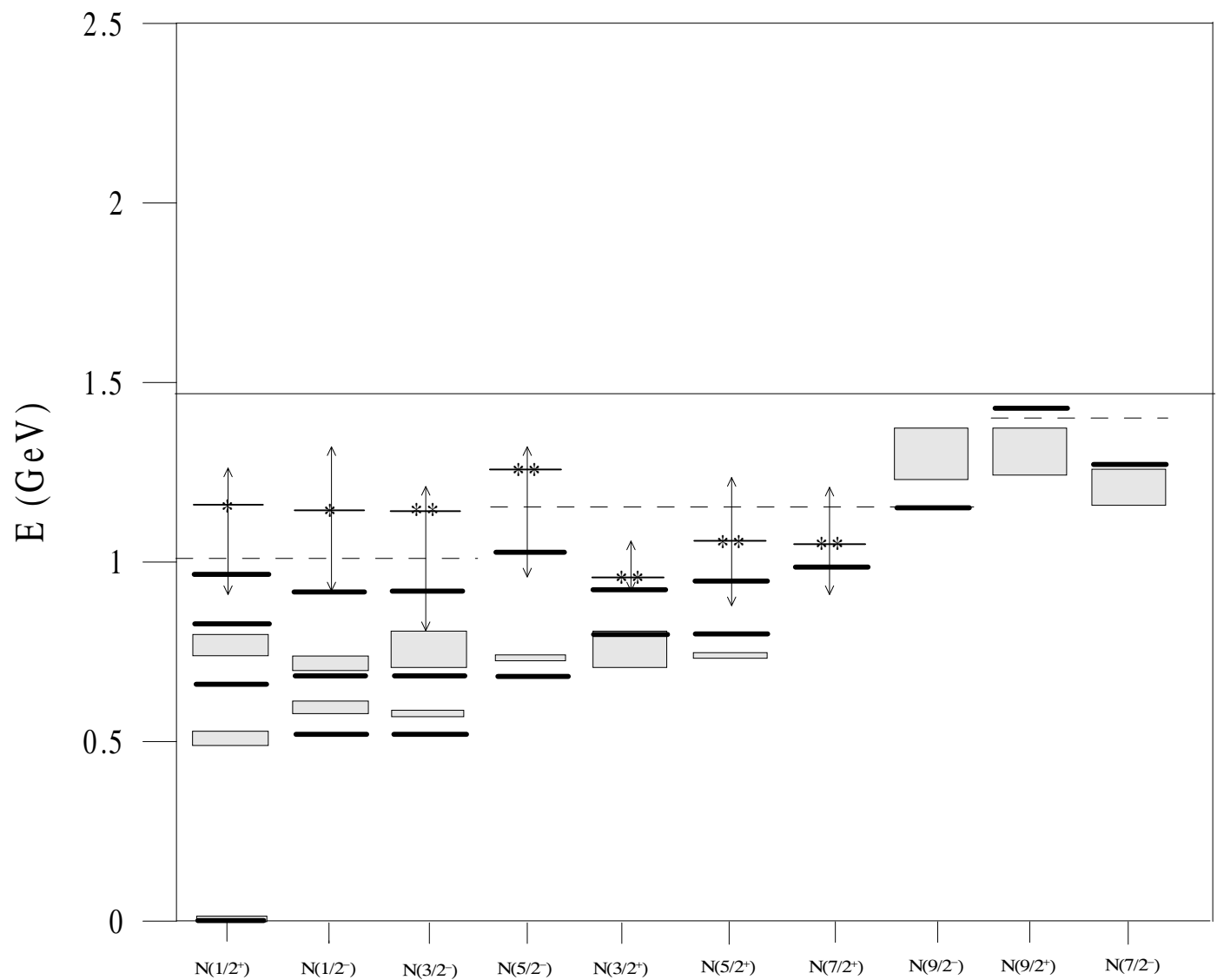
\begin{figure}[tbp]
\caption{ Relative energy nucleon spectrum for the screened potential. The
solid lines represent our results. The shaded region, whose size stands for
the experimental uncertainty, represents the experimental data for those
states cataloged as $(***)$ or $(****)$ states in the Particle Data Book 
\protect\cite{HAGI}. Experimental data cataloged as $(*)$ or $(**)$ states
are shown by solid lines with stars over them and by vertical lines with
arrows standing for the experimental uncertainties. Finally, we show by a
dashed line the $1q$ ionization threshold and by a thin solid line the total
threshold.}
\label{f3}
\end{figure}

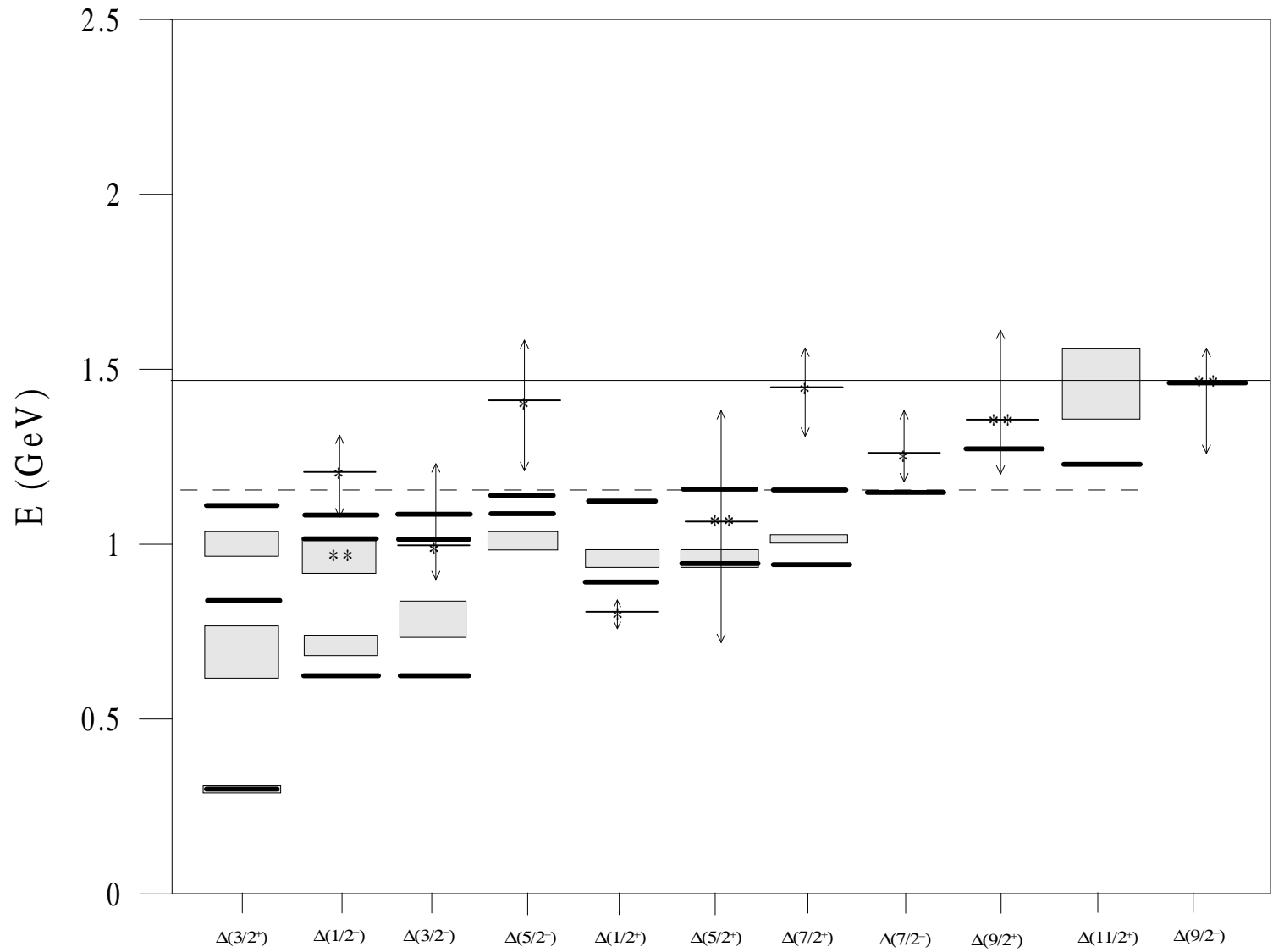
\begin{figure}[tbp]
\caption{ Same as Fig. \ref{f3} for $\Delta$ states.}
\label{f4}
\end{figure}

The predicted spectra for $N$ and $\Delta $ are presented in Figs. \ref{f3}
and \ref{f4} and compared to data. The values of the parameters obtained
after some fine tuning to better accommodate the data appear in Table \ref
{t2}. The $1q$ ionization thresholds for the different values of $(l,s,t)$ are
shown in Table \ref{t3} and plotted in Figs. \ref{f3} and \ref{f4} for the
different $J^{P}$ states.

Let us realize first that at low and medium excitation energies the spectrum
obtained is, according to our hope, of similar or even better quality (due
to the smaller Coulomb strength) than the corresponding to the
nonscreened Bhaduri potential. Regarding the higher excitations the quality
of the description is remarkable (let us note that we have also included the
states we predict close above the thresholds since we do not pretend to
have a precise dynamical description of the thresholds with our simple
model) since apart from keeping the same level of quality than in the low
and medium-lying spectrum we get a perfect (one to one) correspondence
between our predicted states and the experimental (up to 2500 MeV)
resonances for any $J^{P}$. The only {\it exception} appears in the $\Delta
(3/2^{-})$ for which we predict three states and there are two cataloged
resonances by the Particle Data Group. However in our opinion it seems
reasonable that the quite different masses reported in different
measurements for the highest state (1940 and 2057 MeV) may correspond in
fact to two different states. As a consequence if we assume that the
quantitative differences we have with data are due to the lack of a more
complete dynamical treatment that does not change our qualitative result,
the missing resonance problem, as stated in Sec. II, disappears, i. e. there
are no missing resonances that correspond to quark model $3q$ bound states. 
The price to be paid is the appearance of free quark states what
requires a hadronization mechanism to connect them to phenomenology. We
assume that the same pair creation mechanism responsible for screening
induces the hadronization of a ($2q$ state $+1q$ free state) into a
baryon-meson state.

With respect to the values of the parameters our initial expectancies are
confirmed after the fine tuning process to fit the spectrum. For the
spin-spin interaction we have very similar values to the nonscreened case 
and for $1/(2\overline{a}^{2})$ a quite bigger value as can be
checked from Table \ref{t1}. Finally for the Coulomb strength we are
able to reduce its value to approximately one tenth of the nonscreened case
value, what causes an improvement in the description of the radial
excitations. This supposes a big difference between the Coulomb and
spin-spin strengths (they are equal in the non-screened case)
indicating the very effective character of the parameters. It is worthwhile
to mention that if instead we had maintained the nonscreened Coulomb
strength we would have had the possibility to include in our description
some of the highest uncertain states but at the price of loosing quality in
the overall fit to the spectrum.

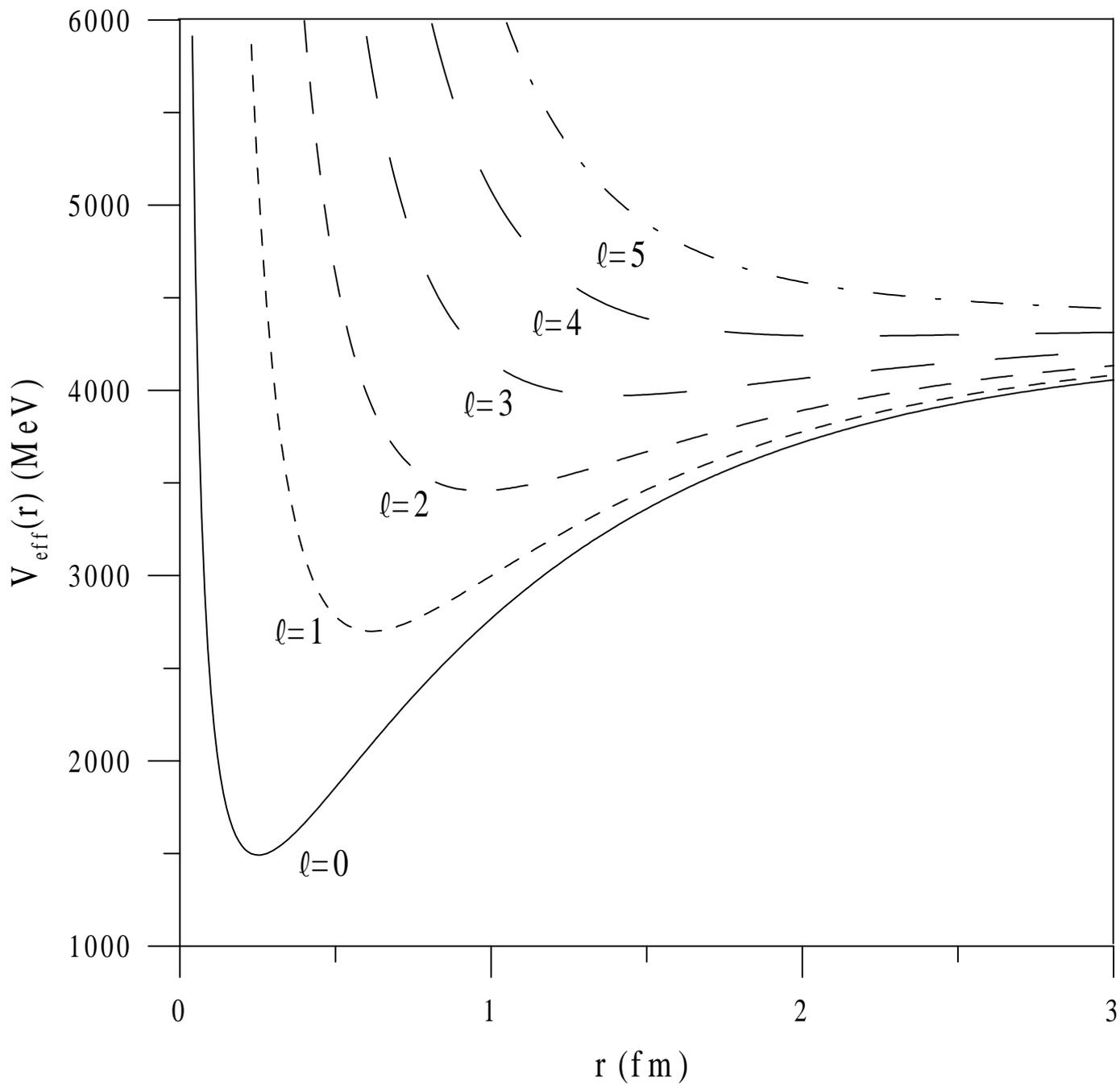
\begin{figure}[tbp]
\caption{ Effective potential as written in Eq. (\ref{efe}) for different
values of $l$, the reduced mass of the light quarks and spin one.}
\label{f2}
\end{figure}

For our choice of the parameters no $3q$ bound state configuration exists
with a minimum value of $l + \lambda >4$. We can understand this result by
drawing in Fig. \ref{f2} an average effective baryonic potential defined as
three times the quark-quark potential plus a centrifugal barrier specified
by orbital angular momentum $l$ or $\lambda $. This is 
\begin{equation}
V_{eff}(r_{ij})=3V_b(r_{ij})+ \frac{{\hbar}^{2}l(l+1)}{2\, m_{red}\,
r_{ij}^{2}}  \label{efe}
\end{equation}
where $m_{red}$ is the reduced mass of a pair of quarks. For this potential
the splitting energy is $3/(2\overline{a}^{2}\mu _{b})$ and no bound state
can be accommodated for $l\geq 5.$ For $l\leq 4,$ apart from the possibility
of having bound states below the $1q$ ionization thresholds, there could
also exist, from the $1q$ ionization thresholds up to the splitting energy,
some metastable $3q$ states as explained above. In our case the
nonconsidered resonances above 2500 MeV could be of this type provided that
their uncertain experimental errors leave opened the possibility for them to
be below the total splitting energy of the system.

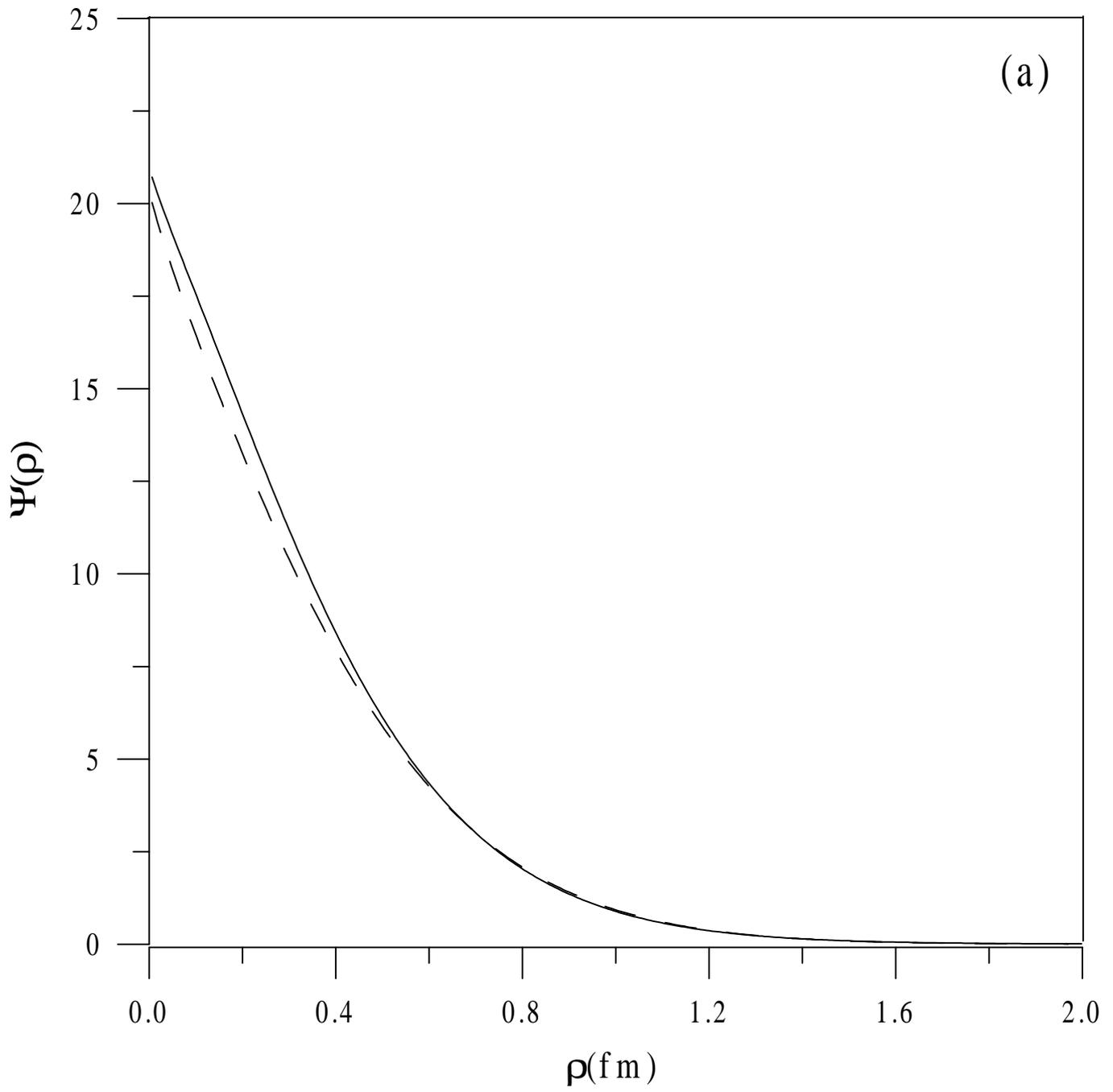
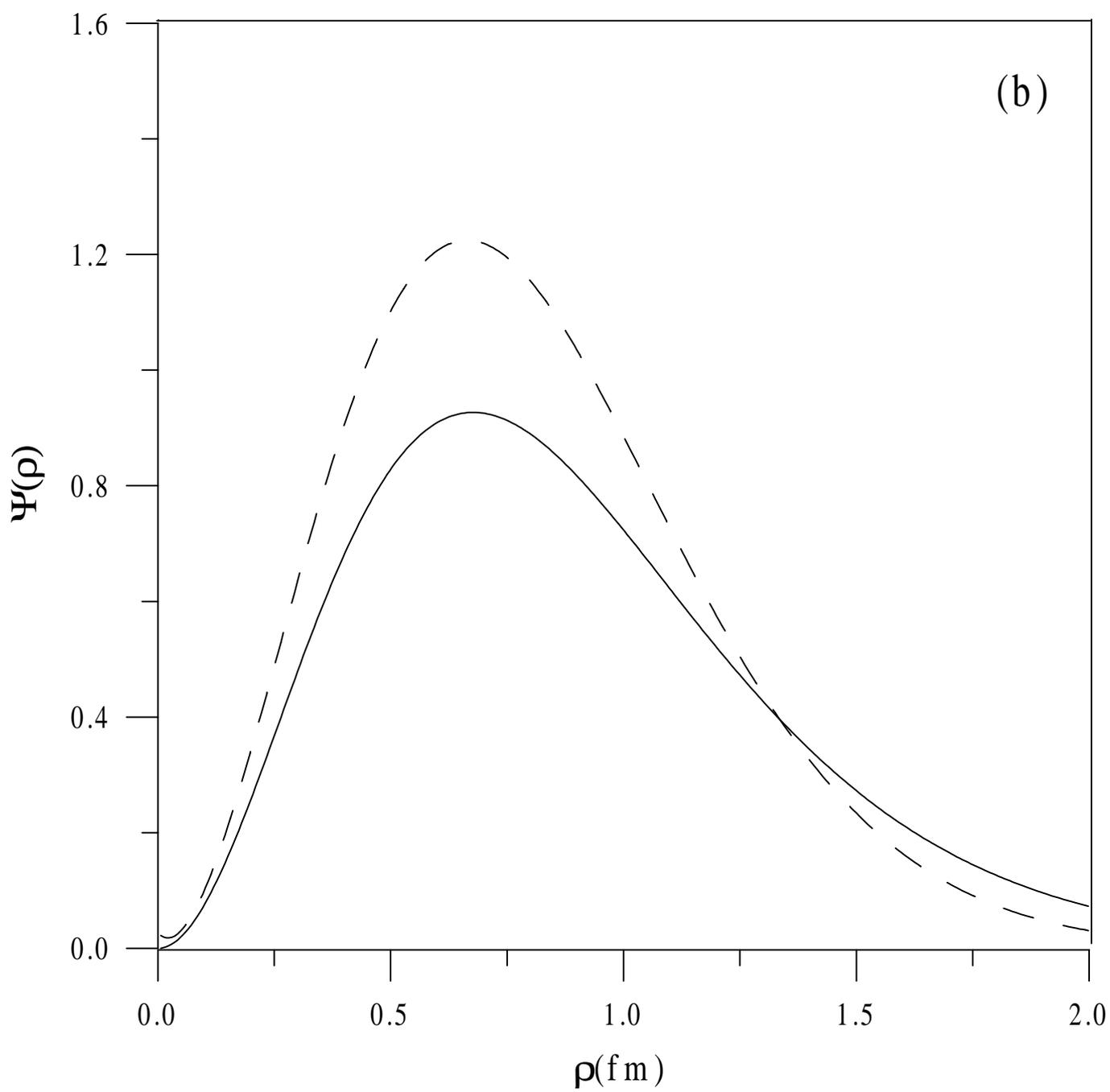
\begin{figure}[tbp]
\caption{ (a) Radial wave function of the completely symmetric component of
the $N(1/2^+)$ in terms of the hyperradius $\protect\rho$. The dotted line
stands for the result of Ref. \protect\cite{BRAC}, the solid line
corresponds to the screened potential. (b) Same as (a) for the dominant
component of the $N(5/2^{+})$ ground state.}
\label{f5}
\end{figure}

It may also be interesting to draw the wave functions. In Fig. \ref{f5}(a)
we plot the ground state nucleon wave functions for the screened and
nonscreened Bhaduri potentials. As we can check the presence of screening
does not mean any major effect but a slight enhancement probability at short
distances compensated by a slight depression at medium range. Just for
comparison we show in Fig. \ref{f5}(b) the wave functions for the ground
state of the dominant component of the $N(5/2^{+})$ where we can appreciate
a more significant difference due to the fact that the large distance part
of the potential is playing a more relevant role.

\section{Summary}

We have examined the consequences, for $N$ and $\Delta $ spectra, of a quark
model description based on a screened potential. The form chosen for the
potential is quite simple containing a linear term plus the 
Coulomb and spin-spin terms of the usual OGE interaction, both
screened through a Yukawa type factor. The OGE parameters, strength for the 
Coulomb and strength and range for the spin-spin terms, are fixed from
the low-lying part of the spectrum whereas the screened linear potential
parameters, strength and range (screening), are determined from the higher
experimental excitations. The resulting average quark velocities inside the
baryon, close to $c$ (one would obtain even bigger values if the
nonrelativistic expression, $p_{q}/m_{q}$, were applied) make clear the
shortcomings of the non-relativistic quark model treatment. 
Nevertheless the description of the spectrum is quite satisfactory.
Except for a few cases [$N(1440),\Delta (1910)$] our results differ at most
100 MeV from data for well established ($\ast \ast \ast $) and ($\ast \ast
\ast \ast $) resonances. Concerning the non-well established resonances it
is also remarkable that below 2500 MeV most of our predictions lie inside
the current experimental error bars. The failure in the description of
particles beyond 2500 MeV may be either an indication of the limit of
applicability of the model, or an indication that we are in the transition
to the continuum including the possibility of metastable states, or a
signature of the presence of exotics (non $3q$ bound states). Anyhow an
experimental effort to clarify the situation in this energy region seems to
be mandatory. Keeping this in mind we can say that the major distinctive
consequence of the use of a screened potential is the finiteness of the
spectrum resulting in an perfect parallelism between predicted $3q$ bound
states and experimental resonances. As a counterpart to this
striking result we are left with a non-confining model. A dynamical
understanding of the decay mechanism to $1q$ free states, involving
hadronization, beyond our qualitative reasonings is mandatory. When this
program is carried out it may well merge out a different view on the
so-called missing resonance problem as an indication that confinement has to
be properly implemented. In this sense we should say that we do not expect
our screening parametrization and our dynamical thresholds for the
existence of $3q$ bound states to be a precise description of the
underlying QCD dynamics (the cross-over from a linear to a flat potential
can be much more rapid in QCD). Despite this the good quality of the high
energy spectrum obtained and its perfect correspondence to data drive us to
think that the qualitative consequences of screening we have derived will be
maintained for more refined potentials. To pursue the effort along this
line, beyond our exploratory work, could help in our opinion to a better
knowledge of the essential ingredients to get an adequate description of
hadrons.

\acknowledgements
We would like to thank very much to V. Vento for his continuous support and
help and to S. Noguera and J. Papavassiliou for suggesting discussions. We
also thank G. Orlandini and W. Leidemann for providing us a code for the HA
method. This work was partially funded by Direcci\'{o}n General de
Investigaci\'{o}n Cient\'{\i}fica y T\'{e}cnica (DGICYT) under the Contract
No. BFM2001-3563, by Junta de Castilla y Le\'{o}n under the Contract No.
SA-109/01, by EC-RTN, Network ESOP, contract HPRN-CT-2000-00130, and by
COFAA-IPN (Mexico).

\begin{table}[tbp]
\caption{Quark model parameters used in Ref. \protect\cite{BRAC}.}
\label{t1}
\begin{tabular}{cccc}
& $m_u=m_d \,\, ({\rm {MeV}) }$ & 337 &  \\ 
& $r_0 \,\, ({\rm {fm}) }$ & 0.45 &  \\ 
& $\kappa \,\, ({\rm {MeV \, fm}) }$ & 102.67 &  \\ 
& $\kappa_\sigma \,\, ({\rm {MeV \, fm}) }$ & 102.67 &  \\ 
& $a \,\, [({\rm {MeV})^{-1}\, {fm}]^{1/2}}$ & 0.0326 & 
\end{tabular}
\end{table}

\begin{table}[tbp]
\caption{Quark model parameters for the calculation of Figs. \ref{f3} and 
\ref{f4}.}
\label{t2}
\begin{tabular}{cccc}
& $m_u=m_d \,\, ({\rm {MeV}) }$ & 337 &  \\ 
& $\overline{r_0} \,\, ({\rm {fm}) }$ & 0.477 &  \\ 
& $\overline{\kappa} \,\, ({\rm {MeV \,fm}) }$ & 10.0 &  \\ 
& $\overline{\kappa_\sigma} \,\, ({\rm {MeV \, fm}) }$ & 120.0 &  \\ 
& $\overline{a} \,\, [({\rm {MeV})^{-1} \, {fm}]^{1/2}}$ & 0.0184 &  \\ 
& $\mu_b \,\, ({\rm {fm}^{-1}) }$ & 1.05 & 
\end{tabular}
\end{table}

\begin{table}[tbp]
\caption{$1q$ ionization thresholds.}
\label{t3}
\begin{tabular}{cccc}
& $(l,s,t)$ & E(MeV) &  \\ 
\tableline & $(0,0,0)$ & 1007 &  \\ 
& $(0,1,1)$ & 1156 &  \\ 
& $(1,0,1)$ & 1388 &  \\ 
& $(1,1,0)$ & 1402 & 
\end{tabular}
\end{table}

\end{document}